\documentclass[galaxies,article,accept,pdftex,moreauthors]{Definitions/mdpi}
\usepackage{ upgreek }
\firstpage{1}
\pubvolume{1}
\issuenum{1}
\articlenumber{0}
\pubyear{2024}
\copyrightyear{2024}
\externaleditor{Academic Editor: Phil Edwards}
\datereceived{26 July 2024}
\daterevised{23 August 2024} 
\dateaccepted{26 August 2024}
\datepublished{ }
\hreflink{https://doi.org/} 

\usepackage[labelformat=simple]{subcaption}

\DeclareCaptionLabelFormat{subcaptionlabel}{\normalfont(\textbf{#2}\normalfont)}
\newcommand\NHXmath{N_{\rm HX}}
\newcommand\BOXmath{\beta_{\rm OX}}

\newcommand\NHImath{N_{\rm HI}}
\newcommand\NHIunit{~{\rm cm}^{-2}}
\newcommand\fluxUnit{~{\rm erg}\times{\rm cm}^{-2}{\rm s}^{-1}}
\newcommand\setSuper[1]{$^{#1}$}

\usepackage{subcaption}
\graphicspath{{./}{figures/}{auto}{Figure}{Figure/auto.python}}%
\usepackage{tipx}

\Title{The Density and Ionization Profile{s} of Optically Dark and High{-}Redshift GRBs Probed by X-ray Absorption} 

\TitleCitation{The Density and Ionization Profile{s} of Optically Dark and
High{-}Redshift GRBs Probed by X-ray Absorption} 

\Author{
  {Eka Puspita Arumaningtyas} 
\setSuper{1,}*\orcidA{}, Hasan Al Rasyid \setSuper{2}\orcidB{}, Maria Giovanna
Dainotti \setSuper{3,4,5}\orcidC{}, Daisuke Yonetoku \setSuper{1}\orcidD{}
}
\AuthorNames{
  Eka Puspita Arumaningtyas, Hasan Al Rasyid, Maria Giovanna
Dainotti, Daisuke Yonetoku
}
\AuthorCitation{
  Arumaningtyas, E.P.; Al Rasyid, H.; Dainotti, M.G.; Yonetoku, D.
}
\address{\setSuper{1} \quad Auth{Faculty of Mathematics and Physics}, Kanazawa {University} 
, Kakumamachi, {Kanazawa}, Auth{920-1192} 
, Ishikawa, Japan; {yonetoku@astro.s.kanazawa-u.ac.jp}
\\
\setSuper{2} \quad Auth{Faculty of Engineering,} Nurul Jadid {University} 
, Paiton, {Probolinggo}, Auth{67291}
, Indonesia; {hasanalrasyid.har@gmail.com} 
\\
\setSuper{3} \quad Division of Science, National Astronomical
Observatory of Japan, Osawa, {Mitaka} 
, {Tokyo}, Auth{181-8588}, Japan; {maria.dainotti@nao.ac.jp}\\
\setSuper{4} \quad The Graduate {University} for Advanced Studies
(SOKENDAI), Osawa, {Mitaka, Tokyo, 181-8588}, Japan\\
\setSuper{5} \quad Space Science Institute, Walnut Street, Boulder, {CO}, 80301,
USA
}

\corres{Correspondence: {ep.arumaningtyas@gmail.com}
; Tel.:
+81-070-4429-4316}

\abstract{The X-ray column density ($N_{\rm HX}$) of gamma-ray bursts
(GRBs) can probe the local environment of their progenitors over a wide
redshift range. Previous work has suggested an increasing trend as a
function of redshift. The relevance of this current analysis relies on
investigating the selection bias method, such as the effect of the X-ray
spectrum in high-redshift GRBs, which complicates the measurement of
small $N_{\rm HX}$; this has yet to be fully evaluated or discussed
elsewhere. In this work, we evaluated these effects through simulations
to define appropriate observational limits in the $N_{\rm HX}$ versus
redshift plane. We then applied a one-sided nonparametric method
developed by Efron and Petrosian. Within the framework of this method,
we investigated the redshift dependence of $N_{\rm HX}$ and the local
distribution function. Our results show that the evolution of
$N_{\rm HX}$ with redshift firmly exists with a significance of more
than four sigma and follows a power law of
\((1+z)^{1.39 (+0.22, -0.27)}\). Based on these analyses and previous
studies, the GRB progenitor mass varies but is more massive in the high-redshift environment and has a higher gas column density. This suggests
that part of the luminosity evolution of GRBs, which has been widely
reported, may be due to the evolution of the progenitor’s mass. Using
the same method, we demonstrate that optically dark GRBs show a
consistent evolution: \((1+z)^{1.15(+0.67, -0.83)}\). By applying the
Kolmogorov--Smirnov (KS) test, it is shown that optically dark GRBs have
statistically identical flux and photon index distributions compared to
normal GRBs, but the $N_{\rm HX}$ is systematically larger. This result
suggests that the darkness of some GRB populations is not due to an
intrinsic mechanism, but rather because a higher density \mbox{surrounds them}.}
\begin{document}
\hypertarget{introduction}{%
\section{Introduction}\label{introduction}}

The measurement of the column density is crucial for identifying the
extragalactic absorption material that accompanies the host galaxy and the
local site of the GRB. The extragalactic absorption provides information
about the local environment of the source and the properties of the
host. In addition, it is important to note that recent work by Dainotti
et al. \citep{dainotti2024, dainotti2024b} has drawn attention to the
importance of the relationship between the neutral hydrogen column
density and the redshift. Their studies emphasize that column density is
the most important predictor of both X-ray and optical data,
highlighting its central role in understanding the properties and
behavior of various \mbox{astrophysical phenomena.}

The Lyman-alpha absorption lines (Ly-\(\alpha\)) in the UV region are a
powerful tool for measuring the hydrogen column density, usually
expressed as $N_{\rm HI}$, in the afterglow of gamma ray bursts (GRBs).
GRBs typically occur in host galaxies with high $N_{\rm HI}$, known as
damped Ly-\(\alpha\) absorbers (DLAs), with
\(\mathrm{log}\ N_{\rm HI}> 20.3\ {\rm cm}^{-2}\) \citep{tanvir2019}.
This is consistent with the massive stellar progenitors of GRBs, which
are known to typically be located within or near gas-rich star-forming
regions such as dense molecular clouds
\citep{jakobsson2006, tanvir2019}.

In the X-ray band, we measure the column density based on the extinction
observed in the low-energy spectrum. This phenomenon is primarily caused
by the photoelectric effect in the medium along the line of sight. In
general, we convert the total column density to an equivalent hydrogen
column density ($N_{\rm HX}$) using the metallicity parameter. This
indicates the total contribution of the different elements to the
absorption process.

Previous research has postulated that the increase in $N_{\rm HX}$ is
closely related to the redshift \citep{rahin2019, dalton2020}. The
observed increase in the $N_{\rm HX}$ trend with increasing redshift has
been attributed to a distortion of dust extinction by optically dark
GRBs \citep{watson2012}. It is noteworthy that this group of GRBs
exhibits a significant proportion (\(20-30\%\)) of the afterglow in the
near-infrared (NIR) range \citep{fynbo2001, greiner2011}.

To classify a GRB as optically dark, two classification methods have
generally been used: the Jakobsson method \citep{jakobsson2004} and the
van der Horst method \citep{vanderhorst2009}. \citet{jakobsson2004}
developed the classification method by comparing X-ray to optical flux
(\(\beta_{\rm OX}\)) for all known GRBs. Using a sample of \(52\) GRBs,
they identified five dark outbursts as outliers with
\(\beta_{\rm OX} < 0.5\). This threshold serves as a criterion for
classifying an outburst as dark using the Jakobsson method. The van der
Horst method uses X-ray flux and spectral information from Swift and
classifies a GRB as dark if its \(\beta_{\rm OX}\) is shallower than
\(\beta_{\rm X} - 0.5\).

Based on the spectral index in the afterglow theory
\citep{sari1998, piran1999} and related references, we expect an energy
spectrum of \(\nu^{-p/2}\) for the high-energy case of slow cooling.
Here, \(p\) denotes the energy index of the accelerated electron. In the
case of \(p=2\), the afterglow spectrum is \(\nu^{-1}\), which leads to
\(\beta_{\rm OX} = 1\). Assuming that the cooling frequency is between
the optical and X-ray band, the spectrum progresses from \(\nu^{-(p-1)/2}\) to
\(\nu^{-p/2}\). Therefore, the expected spectral index is
\(0.5 < \beta_{\rm OX} < 1\).

\textls[-15]{Although the definitions proposed by \citet{jakobsson2004} and 
\citet{vanderhorst2009}} differ, both emphasize that the optical flux is weaker than the expected value based on the extended spectral
index of X-rays described in standard afterglow theory.

The darkness of certain GRB groups has primarily been explained by three
hypotheses. First, an intrinsic luminosity feature of the afterglow
suggests the existence of optically bright and dark subclasses
\citep{lazzati2002}. Second, some GRB groups may be located at high
redshifts, leading to Ly-\(\alpha\) absorption by the intergalactic
medium, which hinders their detection in the R-band at the observer
frame \citep{tanvir2008, cucchiara2011}. Finally, the darkness of
certain GRB groups can be attributed to their location behind very dense
material along the line of sight to the host galaxy
\citep{lazzati2002, perley2009, higgins2019}.

The possibility of an increasing trend of $N_{\rm HX}$ as a function of
redshift has been discussed in several previous papers
\citep{campana2010, behar2011, campana2012, starling2013, rahin2019, dalton2020}.
However, these studies used forward fitting to analyze the data
distribution. In this work, we applied a flux limit to select the
$N_{\rm HX}$ data and then used the Efron and Petrosian method (EP
method) to determine the cosmological evolution of $N_{\rm HX}$.

In contrast to many traditional methods, which assume a certain
distribution of the data, the EP method is nonparametric. This means
that it makes fewer assumptions about the underlying distribution, which
makes it more flexible and robust for different applications. The EP
method was developed specifically for dealing with truncated data, where
only a subset of the data are observable due to selection effects or
other \mbox{limitations \citep{efron1992, efron1999}}. This is common in
astrophysical surveys where only the brightest or closest objects \mbox{are
detectable.}

The EP method enables the estimation of distribution functions and the
investigation of correlations in the presence of truncation. This method
is often used to analyze observational datasets that are subject to
truncation and censoring
\citep{kocevski2006, yonetoku2004, singal2011, dainotti2013, dainotti2015, dainotti2020}.
Researchers such \mbox{as \citet{yu2015}}, \citet{petrosian2015},
\citet{pescalli2016}, \citet{lloyd-ronning2019}, and
\citet{dainotti2024} have used the EP method to determine the luminosity
function \(\Psi(L_{0})\) and the cosmic GRB formation rate \(\rho(z)\).
These studies underline the versatility and robustness of the EP method
in dealing with complex astronomical data.

We have used a method that can calculate the presence or absence of
correlations (in this case, \((1+z)^k\)) in a set of real data
distributions without being affected by the truncation of data
containing unobservable regions (truncation limit) due to the flux limit
and observational bias. It is important to determine a suitable
truncation line. To determine the truncation in the plane
$N_{\rm HX}$ vs. \((1+z)\), not only the flux limit of Swift XRT must be
determined, but also $N_{\rm HX}$, whose accuracy is simultaneously
affected by the observed flux. In this case, we evaluated the flux limit
with a series of spectral simulations using the respond matrix file to
generate fake data and then estimated the accuracy of the $N_{\rm HX}$.

\hypertarget{data-selection-and-analysis-procedure}{%
\section{Data Selection and Analysis
Procedure}\label{data-selection-and-analysis-procedure}}

\hypertarget{data-selection}{%
\subsection{Data selection}\label{data-selection}}

In this study, we analyzed the spectral properties of GRBs from the
SWIFT database. {{We} 
used the 404 $N_{\rm HX}$ values available up
to GRB 240809A, which contained redshift information. We selected the
time-averaged {spectra of photon-counting (PC) mode for the GRBs.} 
It is worth
noting that the influence of flares on the PC mode was found to be
insignificant in most cases \citep{evans2009}. We obtained \(239\) GRBs
with a \(90\%\) confidence level, which means that their
$N_{\rm HX}$ errors were less than \(90\%\).}

Table~\ref{tbl:grbSample} provides detailed information on our GRB data
sample, with each entry accompanied by redshift, $N_{\rm HX}$, and flux
information. We compiled a data set of \(224\) GRBs that met both the
$N_{\rm HX}$ and flux criteria, with flux brighter than
\(10^{-12.25}~{\rm erg}\times{\rm cm}^{-2}{\rm s}^{-1}\), as described
in Section \ref{analysis-procedure}. Within this dataset, we identified \(30\) optically dark
GRBs that met both the $N_{\rm HX}$ and flux limit criteria.

\hypertarget{analysis-procedure}{%
\subsection{Analysis Procedure}\label{analysis-procedure}}

Our study focuses on the X-ray absorption properties of Swift GRBs with
known redshift, including the selection bias in flux-limited data. We
used simulations to set lower limits on detectable $N_{\rm HX}$ values.
We found that small $N_{\rm HX}$ values can only be accurately measured
for bright and low-shifted afterglows. In contrast, for high-shifted
GRBs, we can only accurately measure larger $N_{\rm HX}$ values due to
the redshift effect on the X-ray photon energy, which shifts to lower
energies with increasing redshift by \((1+z)^{1}\). Consequently, large
$N_{\rm HX}$ values can be measured for GRBs with high redshift.

In this study, we used an absorbed power-law function to account for
each power-law decay phase. The absorption was modeled with two
components, namely, phabs and zphabs. The first component was fixed to
the Galactic value, while the second component was fixed to the redshift
of the host galaxy and the assumed solar metallicity. These modeling
strategies allowed us to accurately describe the observed spectra and
extract valuable information from our data.

We generated truncation lines for the $N_{\rm HX}$ redshift plane using
the “fakeit” command in the XSPEC {software}, ver. 12, created by Keith Arnaud, Ben Dorman, and Craig Gordon. This software created in NASA, Greenbelt, MD, USA. The software is accessed online from \url{https://heasarc.gsfc.nasa.gov/lheasoft/} 
. For the simulated spectra,
we assumed a photon index of two for the afterglow spectrum and a
galactic column density of \(2.1 \times 10^{21}\ {\rm cm}^{-2}\), the
average value of all samples. {{The} detection sensitivity of the
SWIFT XRT telescope is
\mbox{\(2\times 10^{-14} ~{\rm erg}\times{\rm cm}^{-2}{\rm s}^{-1}\) in
\(10^{4}\) s} \citep{burrows2005}. To ensure that our simulations are
relevant to our GRBs database, we determined the exposure time as a
fixed parameter, based on the average exposure time from the GRBs
database, which is \(10^{3}\) s.} We then fitted the simulated
spectra with the input model function and extracted the best
$N_{\rm HX}$ value and its error.

A total of \(495,000\) simulations were performed by varying the
redshift (\(1 \leq z \leq 9\)), dividing the X-ray column density into
\(11\) levels (\(10^{20} \leq N_{\rm HX}\leq 10^{22}\)). {{Our}
flux limit exceeds the sensitivity of the SWIFT instrument, which is
\(10^{-14}\) in \(10^4\) s \citep{burrows2005}. However, the
minimum flux in our \(404\) GRBs database is \(10^{-13}\). It is
important to note that observations under the flux limits of
\(10^{-13}\) generally have lower signal-to-noise ratios, which can
decrease the quality of the data.} In this case, we simulated the flux
into \(50\) levels (\(10^{-13} \leq F \leq 10^{-10}\)), with \(100\)
trials for each combination. The cut-off line, shown in the left panel
of Figure~\ref{fig:contourPlot} for each redshift, was determined by
selecting cases where the relative error of $N_{\rm HX}$ was less than
\(0.9\). Later, we created the contour plot based on the difference
between the average $N_{\rm HX}$ value and the value calculated for
different redshift ranges. The flux limit for each $N_{\rm HX}$ value
was determined by finding the cross-section between \(x\) (logarithmic
of flux limit) and \(y\) (X-ray column density) in the region where we
cannot define the contour.

{\begin{figure}[H]
\includegraphics[keepaspectratio=true,width=0.97\linewidth]{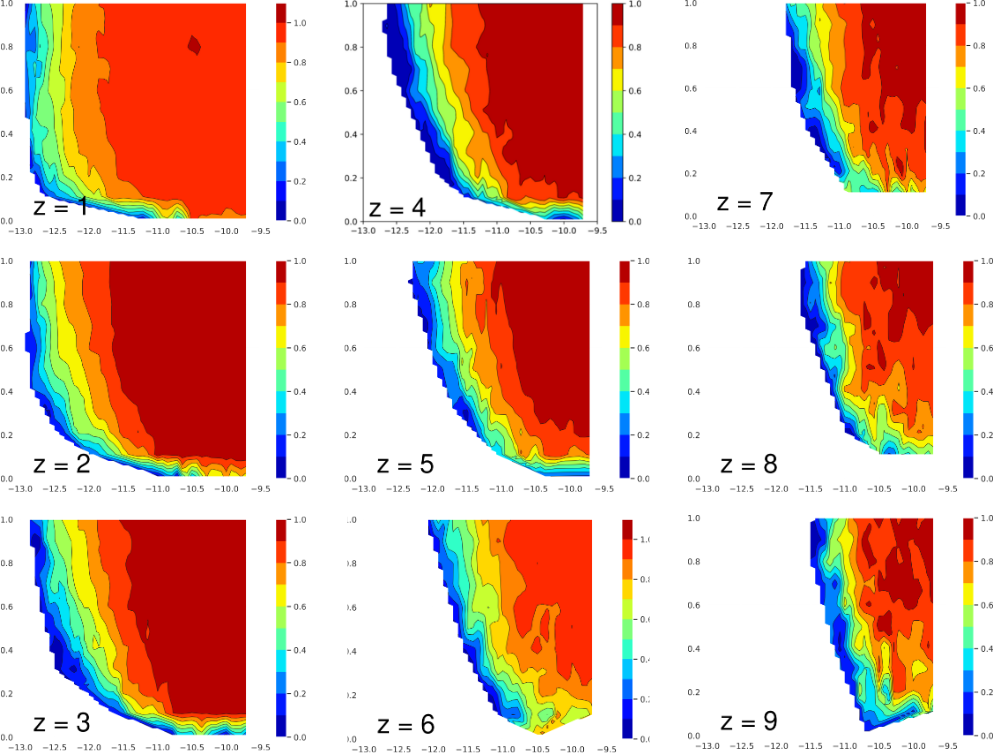}
\caption{
{The} 
 contour plot of the simulated log flux (x-axis) versus
$N_{\rm HX}$ (y-axis) for different \mbox{redshift range}.}
\label{fig:contourPlot}
\end{figure}
}

{{We} analyzed \(404\) GRBs from the Swift database with available
redshift and absorbed flux data. We obtained \(239\) GRBs with a
\(90\%\) confidence level. After applying the flux limit, we obtained
\(224\) data points where both $N_{\rm HX}$ and flux were within the
flux limit. To avoid selecting fewer absorbed GRBs and excluding the most
absorbed ones, we set a high flux limit of
\(10^{-12.25}~{\rm erg}\times{\rm cm}^{-2}{\rm s}^{-1}\), which
preserved a high percentage (\(89\%\)) of the data sample.}

\textls[-15]{The data samples are constrained by truncation, including the flux limit
caused by detector sensitivity. We used a one-sided nonparametric
method developed \mbox{by \citet{dainotti2021}}} to estimate the redshift
evolution of the X-ray column density. This method, derived from
\citet{efron1992,efron1999}, does not involve the correlation of
the parameters. Instead, it evaluates the correlation between
$N_{\rm HX}$ and the redshift. We used the lower bound of the flux limit
for this analysis. By applying this method, we calculated the corrected
distribution of $N_{\rm HX}$.

We compared the cumulative density rate of the optically dark GRBs with
the rest of the GRB data sample of the SWIFT XRT database. We identified
the optically dark GRBs and created a reliable list by applying the flux
limit criteria. We then used the Kolmogorov--Smirnov test to compare the
flux, photon index, and $N_{\rm HX}$ distributions of the two datasets to
determine whether they were from the same population.

\hypertarget{result}{%
\section{Result}\label{result}}

\hypertarget{evolution-of-the-redshift-of}{%
\subsection{\texorpdfstring{Evolution of the redshift of
$N_{\rm HX}$}{Evolution of the redshift of }}\label{evolution-of-the-redshift-of}}

We created a model of the GRB spectrum that includes the absorption of
both galactic and host redshift by using the power-law absorption model
(phabs) and the redshifted power-law absorption model (zphabs).

We used the “fakeit” command in XSPEC to generate simulated spectra,
taking into account parameter ranges for the flux
(\(10^{-13} \leq F \leq 10^{-10}\) with \(50\) steps) and
$N_{\rm HX}$ (\(10^{20} \leq N_{\rm HX}\leq 10^{22}\) with \(11\)
steps). Figure~\ref{fig:contourPlot} shows the contour plot of our
simulated flux versus $N_{\rm HX}$ for different redshift ranges.

{{Figure}~\ref{fig:NHXvsRedshift} (left panel) illustrates an
apparent increase of \(245\) $N_{\rm HX}$ with redshift (including the
outlier). Optically dark GRBs are represented by red triangles, while
others are indicated by black dots. The green crosses represent the
$N_{\rm HX}$ that did not fullfil the flux limit requirement. To
interpret the data distribution accurately, it is crucial to define a
truncation line that is dependent on the flux limit.}

{\begin{figure}[H]
\includegraphics[keepaspectratio=true,width=1.0\linewidth]{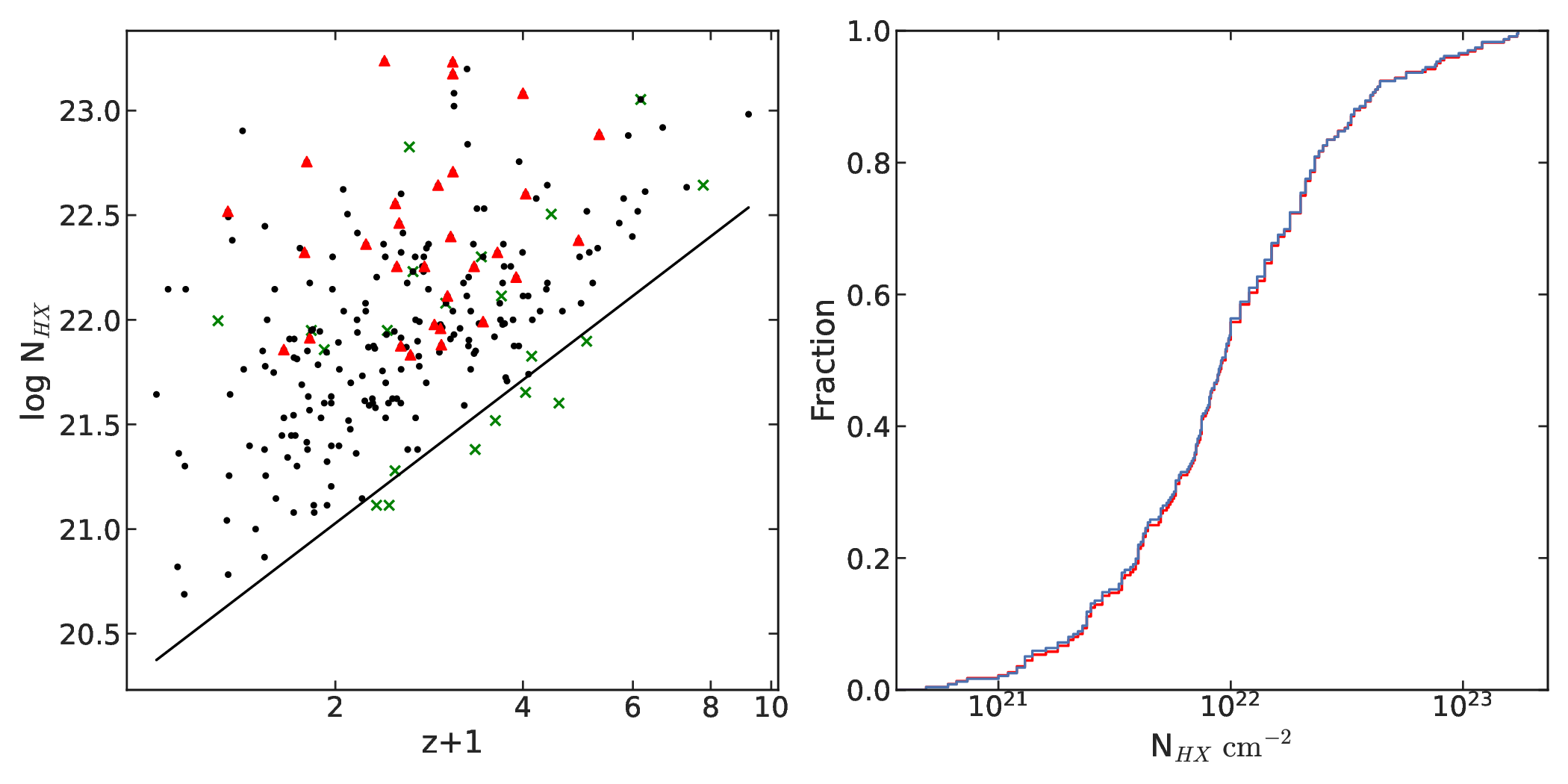}
\caption{
The left panel illustrates the correlation between $N_{\rm HX}$ and the
redshift in our sample of \(245\) GRBs. Optically dark GRBs are
represented by red triangles, while black dots indicate non-dark GRBs.
The green crosses represent data points outside the flux limit. The
solid line represents the assumed sensitivity limit for the column
density given by the flux limit. The right panel shows the cumulative
distribution of GRB data after the removal of certain entries (shown in
blue) and the distribution of remaining entries (shown in red),
resulting in a \emph{p}-value of 0.996}
\label{fig:NHXvsRedshift}
\end{figure}
}

When we exclude data due to the flux limit, it is important that the
remaining data distribution retains the original distribution. We
created cumulative distributions for data selected with different flux
limits and compared them to the original distribution.
Figure~\ref{fig:NHXvsRedshift} (right panel) shows an example of the
$N_{\rm HX}$ distribution for a flux limit of \(10^{-12.25}\), with a
random probability of the KS test of \(0.996\).

We used a one-sided parametric approach developed by Efron and Petrosian
to evaluate the evolution of X-ray column density with redshift in GRB
data \citep{efron1992, efron1999, dainotti2021}. We have assumed that
the evolution of the redshift follows a simple formula, namely,
\(\propto(1+z)^k\). By varying \(k\) from \(0\) to \(2.5\) with a step
size of \(0.1\), we evaluated the randomness of the data distribution
and calculated the test statistic \(\tau\).
Figure~\ref{fig:testStatistic} shows the \(\tau\) values plotted against
the $N_{\rm HX}$ evolution index \(k\) within a three-sigma error range.
We identified the optimal value of \(k\) as \(k_{\rm0}\) at
\(\tau = 0\), which corresponds to an evolution index of
\(k = 1.39^{(+0.22,-0.27)}\) within a one-sigma range. Using the same
method, we determined the evolution properties of optically dark GRBs
and obtained an evolution index of \(k = 1.15^{(+0.67,-0.83)}\) within a
one-sigma range.

{\begin{figure}[H]
\includegraphics[keepaspectratio=true,width=1.0\linewidth]{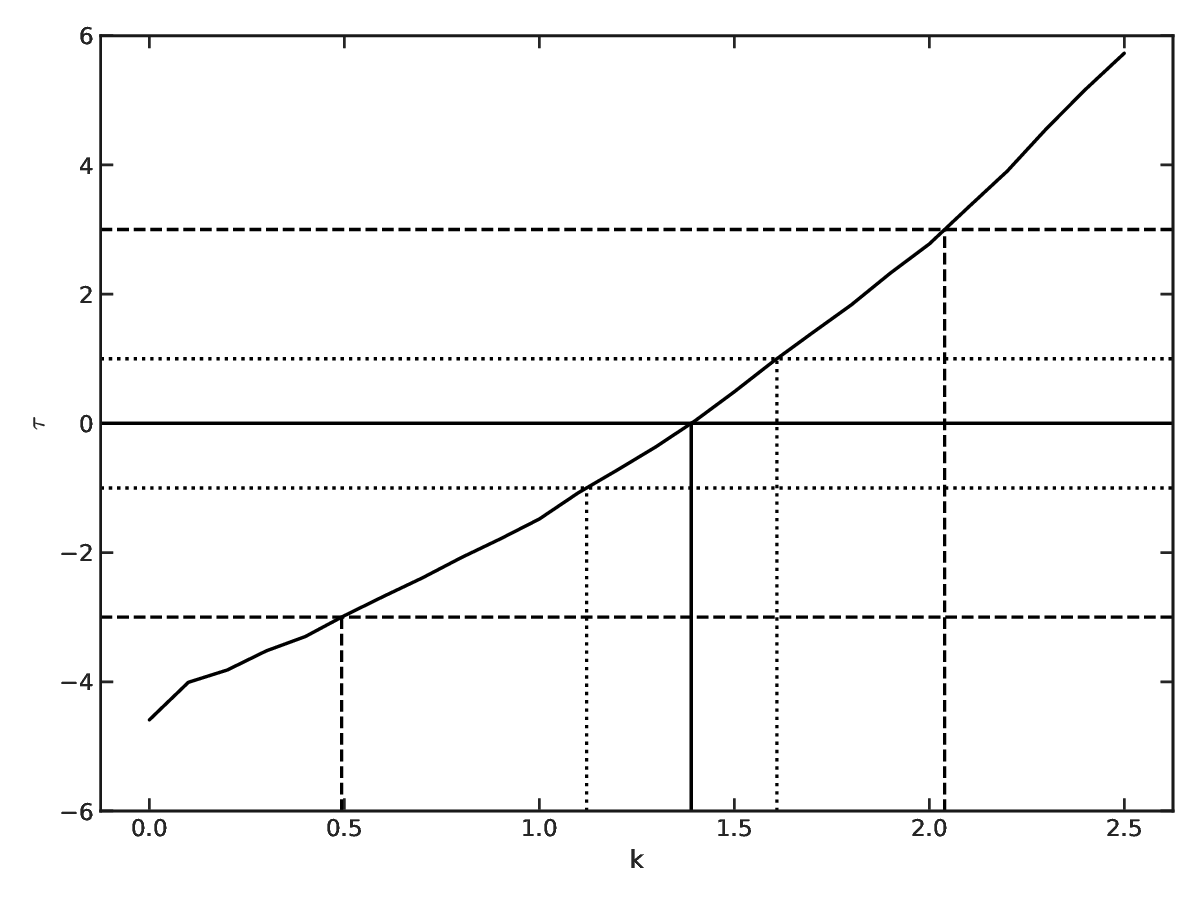}
\caption{
The test statistic \(\tau\) is plotted against the evolution index \(k\)
for the optimal value of \(1.39\). The one-sigma range is from \(1.12\)
to \(1.61\), while the three-sigma range is from \(0.49\) to \(2.04\).}
\label{fig:testStatistic}
\end{figure}
}

\begin{figure}[H]

{\begin{subfigure}{.49\textwidth}
\includegraphics[keepaspectratio=true,width=1.0\linewidth]{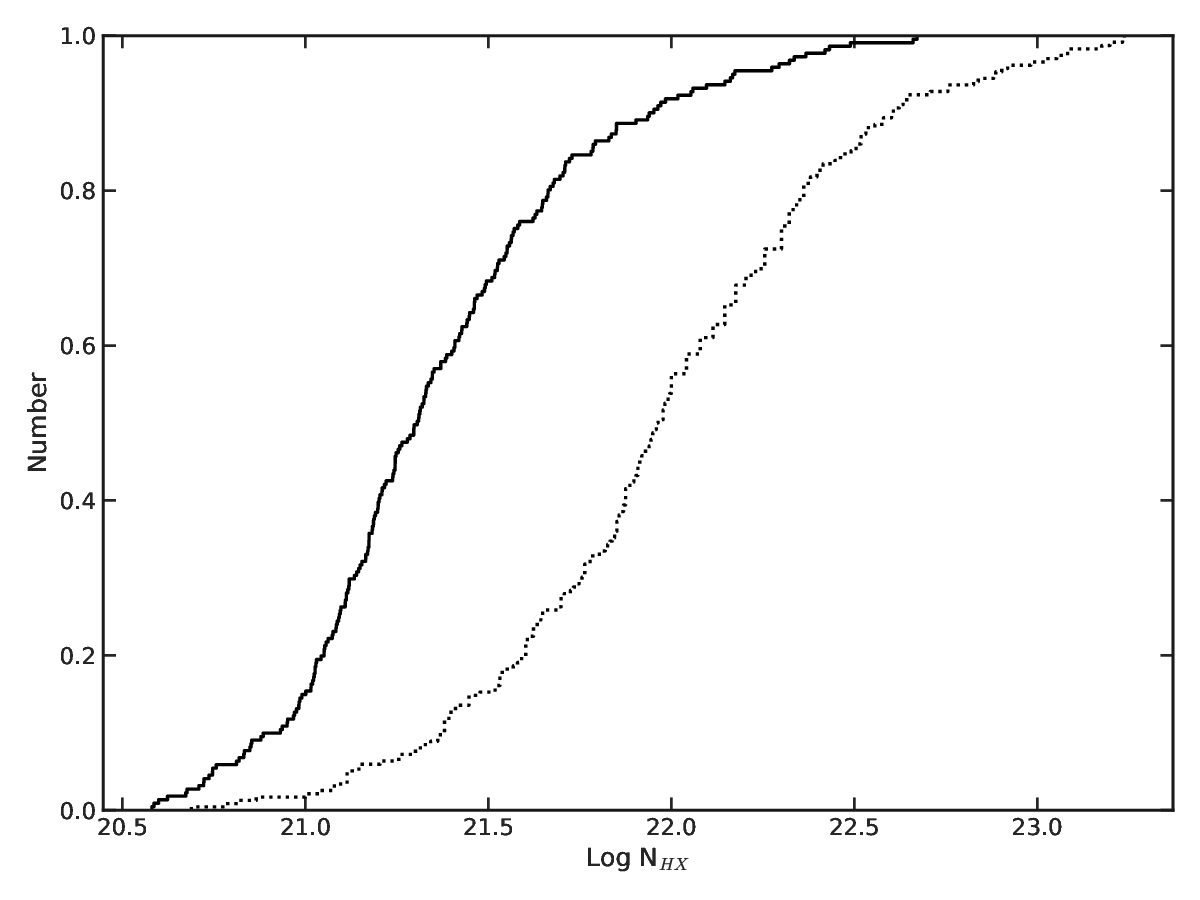}
\caption{
}
\label{fig:cumlogNormal}
\end{subfigure}
}{\begin{subfigure}{.49\textwidth}
\includegraphics[keepaspectratio=true,width=1.0\linewidth]{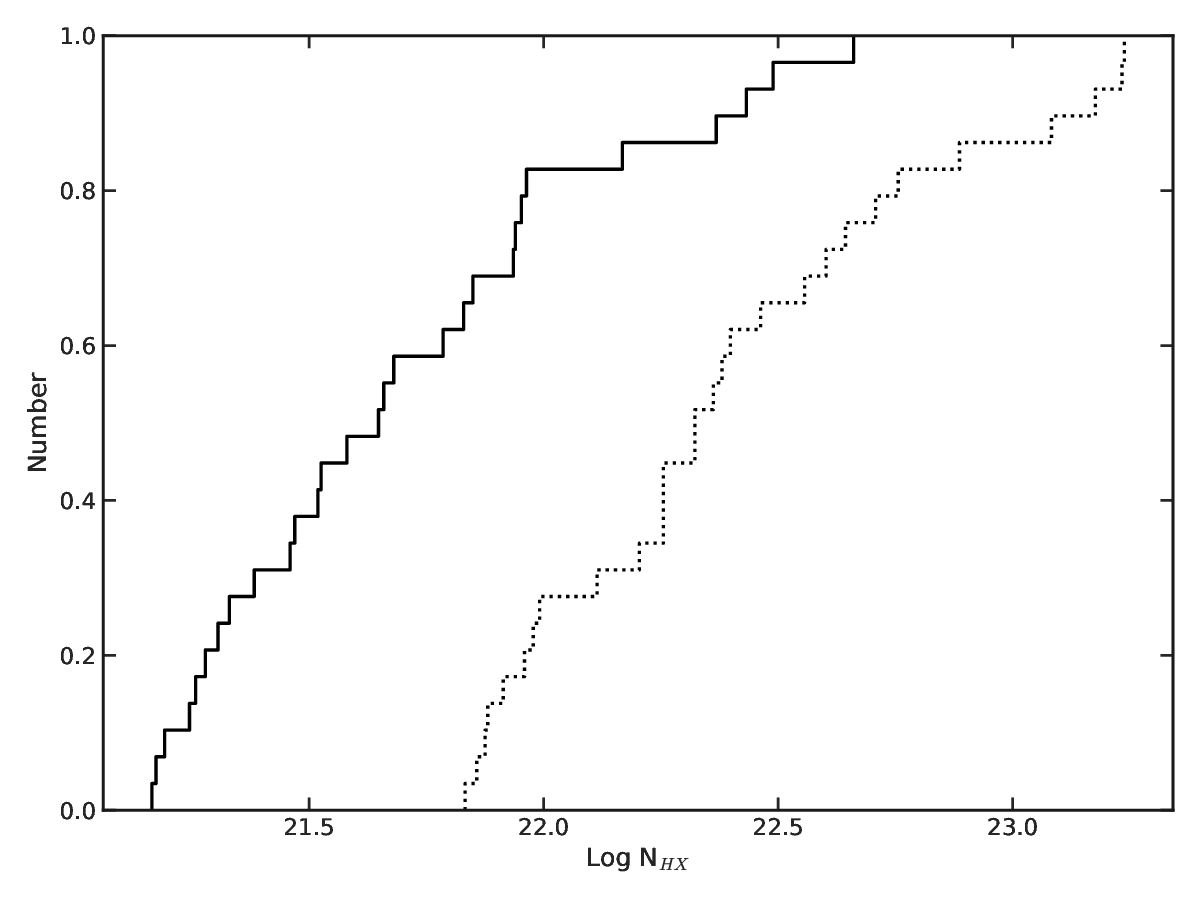}
\caption{
}
\label{fig:cumlogDark}
\end{subfigure}

}

\caption{{ The normalized cumulative distribution of the X-ray column
density ($N_{\rm HX}$) is shown for corrected (solid lines) and
uncorrected (dashed lines) GRBs for normal GRBs, shown in (\textbf{a}), and
optically dark GRBs, shown in (\textbf{b}).}}\label{fig:fig4} \end{figure}

\textls[-15]{Figure~\ref{fig:fig4}a,b illustrate
the distribution of the logarithm of column density (log($N_{\rm HX}$))}
at a redshift of zero \((z=0)\) for normal GRBs and optically dark GRBs.
This distribution, shown as a solid line, has been corrected for
selection biases and takes into account the redshift evolution of the
column density. Using the redshift evolution factor of \((1+z)^{1.39}\)
derived from Figure~\ref{fig:fig4}a, we can extrapolate the
$N_{\rm HX}$ of the normal GRBs distribution at any redshift. Using the
same method, we can determine the $N_{\rm HX}$ distribution of the
optically dark GRB using the redshift evolution factor of
\((1+z)^{1.15}\) derived from Figure~\ref{fig:fig4}b.

\hypertarget{property-of-optically-dark-grb}{%
\subsection{Properties of Optically Dark
GRBs}\label{property-of-optically-dark-grb}}

Figure~\ref{fig:NHXvsRedshift} (left panel) shows that optically dark
GRBs tend to have higher $N_{\rm HX}$ values compared to other GRBs in
the same redshift ranges. This difference is probably due to a dustier
environment. The hosts of optically dark GRBs are characterized by a
higher mass, a redder color, and a higher luminosity
\citep{greiner2011, hjorth2012, chrimes2019}. In addition, the distinct
extinction of the host frame is higher, with \(AV > 1 mag\)
\citep{perley2013}. This evidence supports the conclusion that dust
obscuration is the cause of the optical darkness, which in turn leads to
higher $N_{\rm HX}$ values.

Figure~\ref{fig:NHXvsRedshift} (left panel) shows that GRBs with the
same high $N_{\rm HX}$ can be either bright or optically dark. A
possible explanation for this effect is the variation in the metallicity
of the absorption material, as previously studied by
\citet{chrimes2019}, in host galaxies with higher dust content and solar
metallicity. This idea is also supported by cases such as GRB 020819
\citep{levesque2010} and GRB 051022 \citep{castro-tirado2007}, both of
which exhibit solar metallicity and where no optical nor near-infrared
afterglow was detected. The increased metallicities leading to increased
mass loss could contribute to significant extinction during the lifetime
of the progenitor \citep{trani2014}. These results suggest that
metallicity may play a critical role in determining the observed
properties of GRBs, particularly in X-ray absorption.

\begin{figure}[H]
\centering

{\begin{subfigure}{.49\textwidth}
\includegraphics[keepaspectratio=true,width=1.0\linewidth]{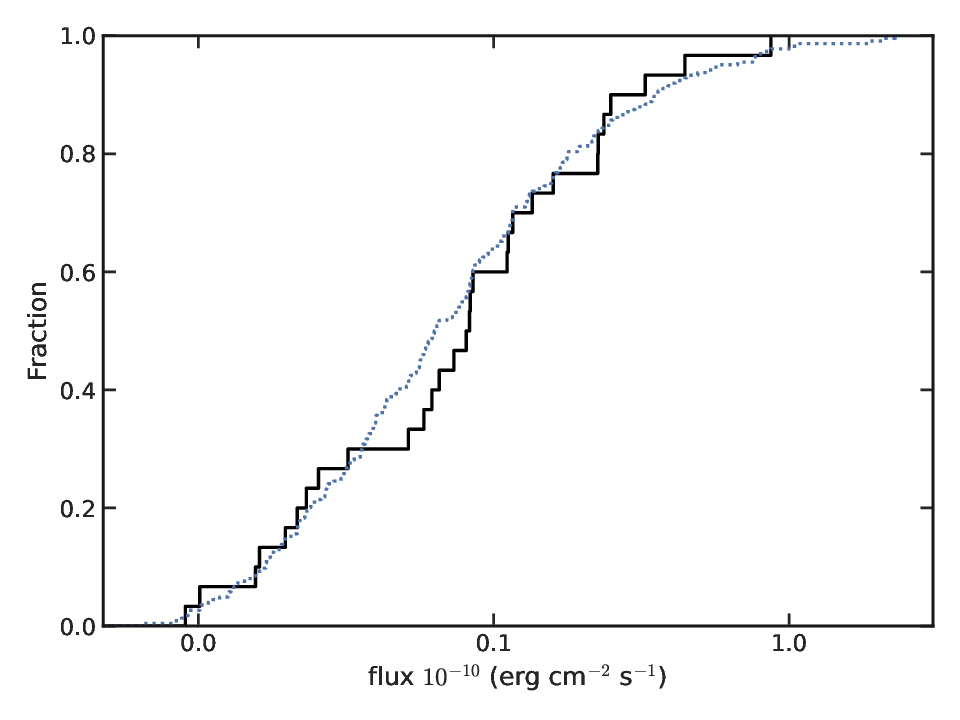}
\caption{
}
\label{fig:ksTestFlux}
\end{subfigure}
\vspace{6pt}
}{\begin{subfigure}{.49\textwidth}
\includegraphics[keepaspectratio=true,width=1.0\linewidth]{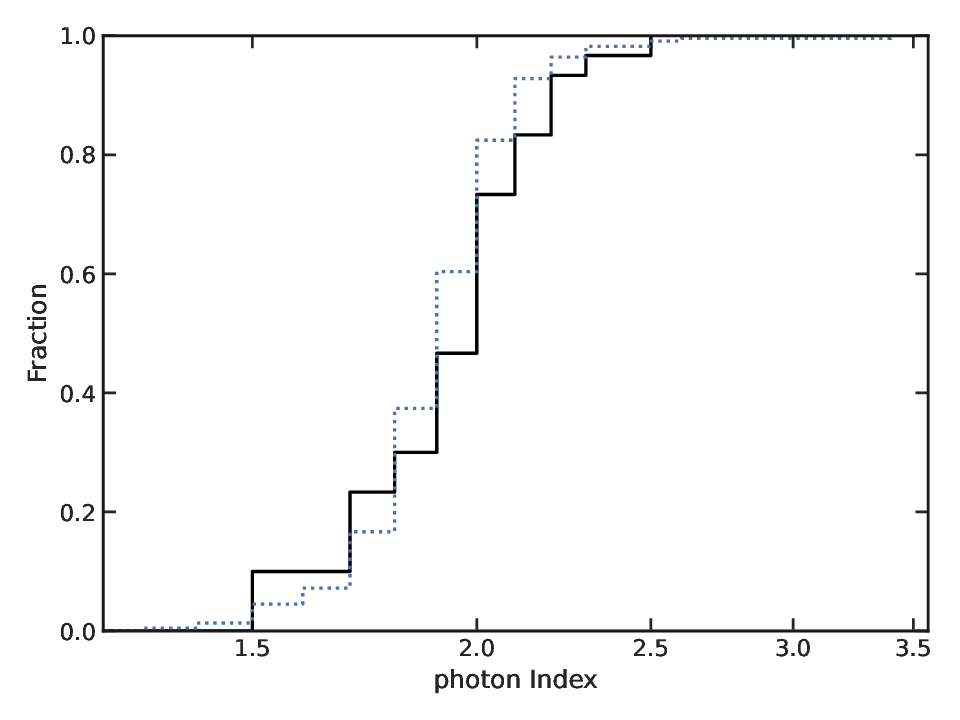}
\caption{
}
\label{fig:ksTestPhoto}
\end{subfigure}
}
{\begin{subfigure}{.49\textwidth}
\includegraphics[keepaspectratio=true,width=1.0\linewidth]{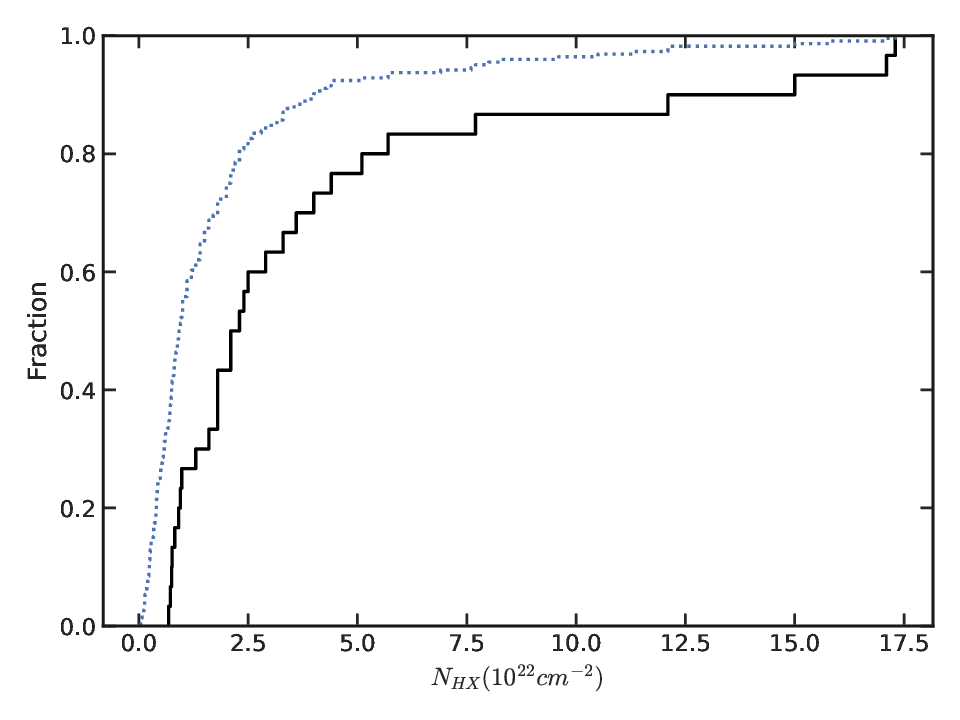}
\caption{
}
\label{fig:ksTestNH}
\end{subfigure}
}

\caption{{ This figure shows the results of the Kolmogorov--Smirnov test
(KS) for two groups: optically dark GRBs (black) and the remaining GRBs
(blue). (\textbf{a}) shows the distribution of the time-averaged flux for the two
groups, giving a KS test statistic of \(0.12\) and a \emph{p}-value of
\(0.76\). In (\textbf{b}), the photon index of the two groups is compared,
resulting in a KS test statistic of \(0.14\) and a \emph{p}-value of \(0.63\).
The distribution of the X-ray column density ($N_{\rm HX}$, right) with
a KS test statistic of \(0.38\) and a \emph{p}-value of \(0.00065\) is represented
in (\textbf{c}).}}\label{fig:fig5} \end{figure}

After applying a flux limit of
\(10^{-12.25}~{\rm erg}\times{\rm cm}^{-2}{\rm s}^{-1}\) to the GRB
data, the results shown in Figure \ref{fig:fig5} emphasize and complement the results
shown in Figure~\ref{fig:NHXvsRedshift} (left panel). Figure \ref{fig:fig5}
represents the results of Kolmogorov--Smirnov tests conducted to compare
three different properties of two groups of gamma ray bursts, optically
dark GRBs and the remaining GRBs, after applying a flux limit of
\(10^{-12.25} ~{\rm erg}\times{\rm cm}^{-2}{\rm s}^{-1}\). First,
Figure \ref{fig:fig5}a displays the distributions of the
time-averaged flux (calculated within \(12\) hours of the outburst) for
both groups. The KS test comparing these distributions yields a
statistic of \(0.12\) and a \emph{p}-value of \(0.76\). This suggests that the
time-averaged flux distributions of optically dark GRBs and the
remaining GRBs are not statistically different. Next,
Figure \ref{fig:fig5}b compares the photon indices of the two
groups. The KS test results in a statistic of \(0.14\) and a \emph{p}-value of
\(0.63\), again indicating no statistically significant difference
between the photon index distributions. Finally,
Figure \ref{fig:fig5}c presents the distributions of the X-ray column
density ($N_{\rm HX}$) for both groups. In this case, the KS test yields
a statistic of \(0.38\) and a \(p\)-value of \(0.00065\). This low
\emph{p}-value indicates a statistically significant difference between the
$N_{\rm HX}$ distributions of optically dark GRBs and the remaining
GRBs.

\hypertarget{the-of-grbs-with-high-redshift}{%
\subsection{\texorpdfstring{The $N_{\rm HX}$ of GRBs with High
Redshift}{The of GRBs with High Redshift}}\label{the-of-grbs-with-high-redshift}}

GRBs with a high redshift generally exhibit a faint afterglow due to
their large distance. If the wavelength of the Ly-\(\alpha\) break at
\(z > 5\) is more than 730 nm, observation in the optical band becomes
difficult. Therefore, its afterglow is typically detected by telescopes
with an aperture of more than \(2\) meters and a near-infrared detector.
The early afterglow can only become visible when these telescopes start
observing and are identified as optically dark GRBs. For example, the
afterglows of the highly displaced GRB 050904 and GRB 90429B were
identified in the near-infrared about \(3\) hours after the outburst
\citep{j.haislip2005}.

Although the number of samples is limited, \(13\%\) of the optically
dark GRBs are from high redshifts, which have a similar environment to
optically dark GRBs at \(z < 5\). Optically dark GRBs account for
20--30\(\%\) of all events, and, of these, \(13\%\), i.e.,~2.6--3.9\(\%\) of
the total, could represent undetected high-redshift events.

Although optically dark and high-shift GRBs share the feature of a high
$N_{\rm HX}$, the origin of the absorption could be different. Due to
the limited energy bandwidth of the Swift XRT (0.3--10 keV), the
measurement of $N_{\rm HX}$ in high-redshift GRBs is a major challenge.

\hypertarget{discussion-and-conclusion}{%
\section{Discussion and Conclusions}\label{discussion-and-conclusion}}

The column density of the GRB population showed a wide range of
characteristics. The relevance of the current analysis lies in its focus
on investigating the selection bias method, particularly in examining the
effects of the X-ray spectrum in high-redshift GRBs. This aspect of the
analysis is crucial because it addresses the challenges associated with
measuring the small values of column density in these distant and energetic
events. The complexities and implications of this measurement difficulty
have not yet been fully evaluated or discussed in other studies, making
this analysis a novel and significant contribution to the field. It is
relevant to note that the work of Dainotti et al.
\citep{dainotti2024, dainotti2024b} recently highlighted the importance
of the relation between column density and redshift, with column
density as the most important predictor for both the X-rays and optical
data. Moreover, in contrast to many other forward-fitting methods used
to solve this problem, we used the Efron--Petrosian method, which allows
a nonparametric determination of these quantities. In this work we have
obtained a precise description of the distribution of these properties---in particular, their cosmological evolution. Our results confirm the
evolution of $N_{\rm HX}$ with the redshift as \((1+z)^{1.39}\). Using
the same method, we obtained a similar trend for optically dark GRBs as
\((1+z)^{1.15}\). Indeed, the fact that we find similar results to
previous studies is not trivial since the relationship between
$N_{\rm HX}$ and the redshift could have been the result of selection
biases rather than intrinsic physics.

{{Our} results on the evolution of $N_{\rm HX}$ indicate that we
might observe either metal content ejected by the progenitor star or a
gas component largely unaffected by cosmic metallicity evolution. Higher
metallicity generally means a greater abundance of heavy elements in the
surrounding medium. Heavier elements are known to absorb X-rays more
efficiently than lighter elements. These results suggest that
metallicity may play a critical role in determining the observed
properties of GRBs, particularly in \mbox{X-ray absorption}.}

{{In} environments with higher metallicity, star clusters
experience greater mass loss compared to those in lower-metallicity
environments \citep{trani2014}. This is because heavier elements in
stellar atmospheres enhance the efficiency of line-driven winds, leading
to more substantial mass loss over time. This process can contribute to
significant extinction in the surrounding medium during the progenitor’s
lifetime, as the expelled material enriches the interstellar medium with
heavy elements and increases the overall density of the surrounding gas
and dust. The increased extinction due to higher metallicity-driven mass
loss can influence the observed properties of the X-ray and optical \mbox{GRB
emission}.}

{{However}, to use metallicity as an explanation for the evolution
of $N_{\rm HX}$, it is important also to consider the evolution of
metallicity itself. For example, host GRBs at redshift \(z < 0.4\) show
a subsolar metallicity cut-off, approximately \(0.4(Z\odot)\)
\citep{niino2017}. Evidence suggests that GRBs at high redshift
preferentially occur in low-metallicity systems, with no clear
indication of an increased metallicity with redshift
\citep{cucchiara2015, graham2023a}. By utilizing a lower metallicity of
\(0.07 \pm 0.05\) \(Z/{Z\odot}\), the calculated X-ray column density is
reduced by {$0.7 - 0.9\%$} \citep{dalton2020}. 
It appears that the
evolution of $N_{\rm HX}$ is not associated with the metals present in
the general interstellar medium of the host galaxy. The variation in
metallicity may help explain why GRBs with the same high
$N_{\rm HX}$ can appear either bright or optically dark, as demonstrated
in Figure~\ref{fig:NHXvsRedshift} (left panel).}

{{In} galaxies with intense star formation, high X-ray column
densities are often attributed to the dense interstellar medium and the
substantial amounts of dust and gas produced by stellar processes. For
GRB hosts, a high star formation rate (SFR) may suggest a young,
actively forming galaxy with significant dust and gas content.
Additionally, high SFRs can be associated with active galactic nuclei
(AGNs) or starburst activity, where intense processes may lead to the
ejection of gas and dust into the surrounding medium. This interaction
can result in higher $N_{\rm HX}$ values, as the burst’s X-ray emission
interacts with the dense material in the host galaxy. Previous studies
have indicated that GRBs exhibit an unusually high formation rate at low
redshifts, exceeding the corresponding SFR by more than an order of
magnitude \citep{dainotti2024, petrosian2024}. However, evidence between
redshifts \(1\) and \(3\) aligns with an almost constant SFR, with a
noticeable decline beyond this range \citep{petrosian2015, dong2022}.
Evidence indicates that those at (\(z > 4\)) are less dusty than those at
lower redshift, with an extinction value of \(A_v < 0.5\) mag
\citep{bolmer2018}. Therefore, the evolution in $N_{\rm HX}$ cannot be
explained by the increasing dust content at higher redshift. The
cosmological evolution of $N_{\rm HX}$ can be attributed to the
accumulation of ionized and highly diffused intergalactic medium (IGM)
\citep{rahin2019}. This intriguing phenomenon needs further
investigation. Despite this, the scatter displayed by $N_{\rm HX}$ poses
a considerable challenge to a straightforward interpretation within the
confines of the IGM model. This complexity suggests that, while the IGM
may contribute to the overall trend, additional factors and mechanisms
may play a role in shaping the observed scatter of $N_{\rm HX}$, for
example, the local interstellar medium (ISM) of the host galaxy. Results
from our simulation suggested that \(13\%\) of GRBs remain in high-density
(\(10^4 ~{\rm cm}^{-3}\)) low-temperature star-forming regions, whereas
87\% of GRBs occur in low-density (\(\sim\) 10--2.5~\({\rm cm}^{-3}\))
high-temperature regions heated by supernovae \citep{cen2014}. A high
percentage of the GRBs remain in a low-density environment, contrasting
with the high $N_{\rm HX}$ and the evolution with redshift. This
contradiction can be explained by certain different models of dust and star
distribution in a very clumpy ISM \citep{corre2018}.}

{{The} stellar initial mass function (IMF) at high redshifts is a
crucial but not yet fully understood aspect of early cosmic events. If
LGRBs originate from massive stars, evidence suggests that the IMF in
the early universe may have been skewed towards larger stellar masses.
Some evidence indicates that the early universe likely favored the
formation of more massive stars than we observe today
\citep{wilkins2008, nanayakkara2017}. Population III stars with masses
ranging from as low as a few to several tens of \(M\odot\)
\citep{stacy2010} to more than a hundred \(M\odot\) \citep{toma2011}
are considered viable progenitors for long-duration GRBs at high
redshift \citep{bromm2006}. Simulation results have found that the
expected fraction of Pop III GRBs is \(\sim10\%\) of the full GRB
population at \(z > 6\) and becomes as high as \(40\%\) at \(z > 10\)
\citep{campisi2011}.}

{{GRBs} are detectable at very high redshifts, potentially beyond
\(z\approx 10\) \citep{lamb2000}, and especially now with the new
observations of high-z galaxies observed at \(z=14\), which can host
population III stars. Their observed properties can provide direct
information about the universe’s conditions during the epoch of
reionization (EoR), which occurred roughly between \(z\approx 6z\) and
\(z \approx 10z\). Only if the average {escape fraction is sufficiently
high, i.e.,~}\(fesc\) {at least 0.1--0.2}
\citep{planckcollaboration2016}, {it is probable} that this phase change
was primarily caused by EUV starlight.} 

{{Following} \citet{chen2007}, the average escape fraction is
inversely proportional to the column density. The total escape fraction
inferred from the GRB column density is one magnitude lower than the
expected value \citep{tanvir2019}. There is evidence from simulations
that GRBs with the same massive progenitor can remain in the
star-forming region. Simultaneously, another population may likely
migrate to a less dense \mbox{environment \citep{cen2014}}. This could occur
due to significant relative motions between stars and their birth
clouds. Numerical models suggest that a relative motion of
\(10~{\rm km s}^{-1}\) between a star and its birth cloud could result
in a displacement of \(100 {\rm pc}\) over a lifetime of 10 Myr. Runaway OB stars
typically exhibit velocities relative to their birth clouds of around
\mbox{\(20~{\rm km s}^{-1}\) \citep{phillips2024}}. Consequently, runaway OB stars
have played a significant role in displacing GRBs from their birth
clouds. The presence of GRBs in hot, low-density environments is also
influenced by supernova heating from earlier supernovae within the birth
clouds. These two effects contribute almost equally to placing most GRBs
in low-density, high-temperature regions \citep{cen2014}.
\citet{lefloch2012} demonstrate that GRB 980425, which occurred in a
nearby (\(z = 0.0085\)) SBc-type dwarf galaxy, appears to be displaced
by \mbox{0.9 kpc} from the nearest HII region. The evolution of
$N_{\rm HX}$ suggests that there is a population of GRBs with low column
density (less than log \(19\)) that have been missed \mbox{in detections}.}

{{The} implication of undetected optically dark GRBs can lead to a
misinterpretation of the GRB’s host galaxies metallicity. The
metallicity distribution derived from analyzing optical afterglow
spectroscopy might not fully represent the true distribution of
metallicities found in the entire universe. This potential discrepancy
arises because the method relies heavily on the observations of optical
afterglows. However, there may be a bias in this observational approach:
it tends to favor detecting afterglows from sources that exhibit lower
metallicities, while bursts associated with higher metallicities might
be less visible or even undetectable in optical wavelengths. This
situation can occur because some bursts, particularly those with higher
metallicity, might be optically dark GRBs. Consequently, the observed
distribution of metallicities based on these optical afterglows could
underrepresent the actual range of metallicities present, especially if
many high-metallicity bursts are not identified or are missed
altogether. This limitation underscores the need for complementary
observational strategies or technologies to capture a more accurate and
comprehensive picture of the underlying metallicity distribution in the
universe. THESEUS will identify more than \(20\) events at \(z > 6\) in
3.5 years of nominal operation, thus providing a much larger sample of
high-redshift GRBs than achieved so far \citep{tanvir2021}. High-z
Gundam mission might be comparable to THESEUS (5 event per year)
\citep{yonetoku2023}. SVOM is already operational, equipped with the 4--
150 keV wide-field trigger camera and has sensitivity to a wide variety
of GRBs, including those at high redshifts. It is expected to detect
around 80--100 GRBs per year \citep{lanza2024}.}

{{In} summary, we used simulations to establish observational flux
limits and applied a one-sided nonparametric method by Efron and
Petrosian to understand the correlation between the $N_{\rm HX}$ and
redshift. We found a significant evolution of $N_{\rm HX}$ with
redshift, following a power law of \((1+z)^{1.39 (+0.22, -0.27)}\). For
optically dark GRBs, a similar evolution of
\((1+z)^{1.15^(+0.67, -0.83)}\) was obtained. A Kolmogorov--Smirnov test
showed that optically dark GRBs have similar flux and photon index
distributions as normal GRBs, but a larger $N_{\rm HX}$, indicating that
their darkness results from a higher surrounding density rather than an
intrinsic mechanism. The evolution of $N_{\rm HX}$ suggests that GRB
progenitors are more massive and have higher gas column densities at
higher redshifts, potentially contributing to GRB luminosity evolution.}

{


\begin{table}[H]
\tablesize{\footnotesize}

\caption{\label{tbl:grbSample}{GRB data} 
}


\begin{adjustwidth}{-\extralength}{0cm}
\begin{tabularx}{\fulllength}{lCCccccc}
\toprule
\multirow{2}{*}{\textbf{GRB Name}} & \multirow{2}{*}{\textbf{Redshift}} & \boldmath{\(\NHXmath\)} & \boldmath{\(\NHXmath\)} \textbf{Error} & \multirow{2}{*}{\boldmath{\(\BOXmath\)}} &
\boldmath{\(\NHImath\)} & \textbf{Flux} & \textbf{Flux Error}\\
& & \boldmath{\(~10^{22}~{\rm cm}^{-2}\)} & \boldmath{\(~10^{22}~{\rm cm}^{-2}\)} & &
\boldmath{\(~10^{22}\NHIunit\)} & \boldmath{\(~10^{-12}~\fluxUnit\)} & \boldmath{\(~10^{-12}~\fluxUnit\)}
\\
\midrule


GRB050315 & 1.95 & 0.95 & 0.46 & 0.63 \(^{3}\) & & 9.50 & 0.45\\
GRB050401{* } 
\(^{3,a}\) & 2.90 & 1.60 & 1.94 & 0.36 \(^{3}\) & 3.98
\(^{3,5}\) & 22.60 & 1.96\\
GRB050408 & 1.24 & 1.20 & 0.48 & & & 3.78 & 0.30\\
GRB050416A* & 0.65 & 0.72 & 0.24 & 0.35 \(^{6}\) & & 1.56 & 0.09\\
GRB050505 & 4.27 & 2.20 & 1.34 & 0.53 \(^{3}\) & 1.26 \(^{5}\) & 3.04 &
0.12\\
GRB050525A & 0.61 & 0.14 & 0.24 & 0.92 \(^{3}\) & & 5.26 & 0.48\\
GRB050802 & 1.71 & 0.24 & 0.24 & 0.51 \(^{3}\) & & 5.22 & 0.21\\
GRB050803* & 4.30 & 7.70 & 3.00 & $-$0.15 \(^{3}\) & & 6.17 & 0.29\\
GRB050822 & 1.43 & 0.40 & 0.32 & & & 3.98 & 0.26\\
GRB050922B & 4.90 & 7.60 & 3.40 & 0.58 \(^{3}\) & & 67.00 & 3.70\\
GRB051006 & 1.06 & 4.20 & 2.60 & 1.30 \(^{3}\) & & 1.02 & 0.13\\
GRB051016B & 0.94 & 0.70 & 0.38 & 0.63 \(^{3}\) & & 1.53 & 0.12\\
GRB051022* & 0.80 & 5.70 & 1.28 & $-$0.10 \(^{2}\) & & 15.90 & 0.82\\
GRB051111 & 1.55 & 0.58 & 0.66 & & & 1.29 & 0.12\\
GRB051221A & 0.55 & 0.18 & 0.18 & & & 1.70 & 0.13\\
GRB060124 & 2.30 & 0.58 & 0.56 & 0.80 \(^{3}\) & 0.07 \(^{5}\) & 8.55 &
0.32\\
GRB060202* & 0.79 & 2.10 & 0.52 & 0.20 \(^{3}\) & & 1.61 & 0.09\\
GRB060204B* & 2.34 & 1.80 & 0.84 & 0.47 \(^{3}\) & & 2.32 & 0.14\\
GRB060210* & 3.91 & 2.40 & 1.22 & 0.37 \(^{3}\) & 0.35 \(^{3,5}\) &
11.10 & 0.35\\
GRB060218 & 0.03 & 0.44 & 0.12 & & & 0.93 & 0.05\\
GRB060306 & 1.55 & 4.00 & 1.20 & 0.54 \(^{3}\) & & 1.91 & 0.12\\
GRB060502A & 1.51 & 0.42 & 0.44 & 0.53 \(^{6}\) & & 8.79 & 0.66\\
GRB060522 & 5.11 & 3.30 & 5.80 & 0.74 \(^{3}\) & 0.01 \(^{5}\) & 2.24 &
0.23\\
GRB060604 & 2.68 & 1.00 & 1.06 & 0.75 \(^{3}\) & & 1.32 & 0.09\\
GRB060607A & 3.08 & 0.55 & 0.80 & 0.53 \(^{3}\) & 0.10 \(^{5}\) & 17.90
& 0.76\\
GRB060714 & 2.71 & 0.95 & 1.18 & 0.77 \(^{3}\) & 0.71 \(^{5}\) & 8.40 &
0.57\\
GRB060719* & 1.53 & 2.90 & 1.04 & $-$0.13 \(^{3}\) & & 2.55 & 0.18\\
GRB060729 & 0.54 & 0.07 & 0.04 & 0.80 \(^{3}\) & & 10.70 & 0.25\\
GRB060814* & 1.92 & 4.40 & 0.98 & $-$0.06 \(^{3}\) & & 5.14 & 0.22\\
GRB060904B & 0.70 & 0.28 & 0.32 & & & 2.43 & 0.20\\
GRB060908* & 1.88 & 0.95 & 0.92 & 0.38 \(^{3}\) & & 8.28 & 0.86\\
GRB060912A & 0.94 & 0.21 & 0.32 & 0.62 \(^{3}\) & & 4.29 & 0.42\\
GRB060926 & 3.20 & 3.80 & 5.80 & & 3.16 \(^{5}\) & 1.08 & 0.18\\
GRB061007 & 1.26 & 0.74 & 0.32 & 0.79 \(^{3}\) & & 8.62 & 0.50\\
GRB061021 & 0.35 & 0.06 & 0.05 & 0.75 \(^{3}\) & & 3.89 & 0.14\\
GRB061121 & 1.31 & 0.73 & 0.24 & 0.64 \(^{3}\) & & 15.60 & 0.61\\
GRB061202 & 2.25 & 15.80 & 4.00 & & & 15.90 & 0.78\\
GRB061222A* & 2.09 & 5.10 & 1.24 & $-$0.19 \(^{6}\) & & 22.50 & 0.85  \\
GRB070125 & 1.55 & 0.40 & 0.70 & & & 2.60 & 0.30\\
GRB070129 & 2.34 & 0.69 & 0.84 & 0.62 \(^{3}\) & & 2.75 & 0.17\\
GRB070208 & 1.17 & 0.87 & 0.58 & 0.54 \(^{6}\) & & 1.47 & 0.16\\
GRB070306* & 1.50 & 3.60 & 0.88 & 0.23 \(^{3}\) & & 11.60 & 0.54\\
GRB070318 & 0.84 & 0.90 & 0.28 & 0.78 \(^{3}\) & & 2.72 & 0.16\\
GRB070328* & 2.06 & 2.50 & 0.70 & 0.31 \(^{3}\) & & 8.51 & 0.37\\
GRB070419A & 0.97 & 0.43 & 0.66 & 0.87 \(^{6}\) & & 2.33 & 0.43\\
GRB070419B* & 1.96 & 0.76 & 0.56 & 0.25 \(^{3}\) & & 44.40 & 1.91\\
GRB070508* \(^{3,a}\) & 0.82 & 0.82 & 0.46 & & & 11.20 & 0.72\\
GRB070521* & 2.09 & 15.00 & 4.20 & $-$0.06 \(^{3}\) & & 13.50 & 0.83 \\
GRB070714B & 0.92 & 0.40 & 0.38 & & & 0.92 & 0.08\\
GRB070721B & 3.63 & 1.10 & 1.68 & 0.72 \(^{3}\) & 0.32 \(^{5}\) & 21.50
& 1.34\\
GRB070724A & 0.46 & 0.25 & 0.40 & & & 0.88 & 0.16\\
GRB070810A & 2.17 & 0.91 & 0.90 & & 1.41 \(^{5}\) & 1.93 & 0.17\\
GRB071010A & 0.98 & 2.00 & 1.78 & 0.89 \(^{6}\) & & 0.82 & 0.13\\
GRB071021 & 2.45 & 2.00 & 1.28 & & & 1.75 & 0.13\\

\bottomrule
\end{tabularx}

\end{adjustwidth}
\end{table}

\begin{table}[H]\ContinuedFloat
\tablesize{\footnotesize}

   \caption{\textit{Cont}.}


\begin{adjustwidth}{-\extralength}{0cm}
\begin{tabularx}{\fulllength}{lCCccccc}
\toprule
\multirow{2}{*}{\textbf{GRB Name}} & \multirow{2}{*}{\textbf{Redshift}} & \boldmath{\(\NHXmath\)} & \boldmath{\(\NHXmath\)} \textbf{Error} & \multirow{2}{*}{\boldmath{\(\BOXmath\)}} &
\boldmath{\(\NHImath\)} & \textbf{Flux} & \textbf{Flux Error}\\
& & \boldmath{\(~10^{22}~{\rm cm}^{-2}\)} & \boldmath{\(~10^{22}~{\rm cm}^{-2}\)} & &
\boldmath{\(~10^{22}\NHIunit\)} & \boldmath{\(~10^{-12}~\fluxUnit\)} & \boldmath{\(~10^{-12}~\fluxUnit\)}
\\
\midrule

GRB071025 & 4.80 & 3.80 & 3.00 & 0.50 \(^{3}\) & & 3.74 & 0.21\\
GRB071117 & 1.33 & 1.60 & 0.90 & 0.58 \(^{3}\) & & 1.67 & 0.17\\
GRB080205 & 2.72 & 2.30 & 2.60 & 0.79 \(^{3}\) & & 1.34 & 0.13\\
GRB080207* \(^{2,a}\) & 2.09 & 17.10 & 5.00 & 0.30 \(^{2}\) & & 1.97 &
0.15\\
GRB080210* \(^{1,b}\) & 2.64 & 2.10 & 2.00 & & 0.79 \(^{3,5}\) & 0.90 &
0.09\\
GRB080319A* & 2.03 & 1.30 & 1.12 & 0.41 \(^{6}\) & & 5.80 & 0.65\\
GRB080319B & 0.94 & 0.13 & 0.11 & 0.52 \(^{6}\) & & 34.40 & 1.42\\
GRB080319C* \(^{3,a}\) & 1.95 & 0.91 & 0.68 & 0.36 \(^{6}\) & & 86.80 &
5.99\\
GRB080325* \(^{5,a}\) & 1.78 & 1.80 & 1.70 & 0.33 \(^{4}\) & & 1.01 &
0.14\\
GRB080411 & 1.03 & 0.58 & 0.14 & & & 20.80 & 0.59\\
GRB080413B & 1.10 & 0.33 & 0.17 & & & 5.15 & 0.22\\
GRB080430 & 0.77 & 0.49 & 0.16 & & & 3.60 & 0.18\\
GRB080520 & 1.55 & 2.10 & 2.20 & & & 1.26 & 0.21\\
GRB080602 & 1.82 & 1.40 & 0.74 & & & 186.00 & 12.30\\
GRB080603A & 1.69 & 1.00 & 1.00 & & & 3.26 & 0.38\\
GRB080605* \(^{3,a}\) & 1.64 & 0.68 & 0.74 & & & 8.07 & 0.63\\
GRB080607* \(^{3,a}\) & 3.04 & 4.00 & 1.50 & & 5.01 \(^{3,5}\) & 7.33 &
0.43\\
GRB080707 & 1.23 & 0.41 & 0.60 & & & 1.27 & 0.14\\
GRB080710 & 0.85 & 0.12 & 0.19 & & & 2.68 & 0.19\\
GRB080721 & 2.60 & 0.83 & 0.62 & & 0.20 \(^{5}\) & 12.90 & 0.47\\
GRB080805* \(^{1,a,b}\) & 1.51 & 1.80 & 1.52 & & & 2.16 & 0.29\\
GRB080905A & 0.12 & 0.23 & 0.38 & & & 12.80 & 2.17\\
GRB080905B & 2.37 & 3.40 & 1.42 & & 0.00 \(^{5}\) & 7.80 & 0.48\\
GRB080916A & 0.69 & 0.81 & 0.36 & & & 5.83 & 0.49\\
GRB080928 & 1.69 & 0.34 & 0.42 & & & 5.62 & 0.34\\
GRB081007 & 0.53 & 0.71 & 0.26 & & & 5.47 & 0.43\\
GRB081109 & 0.98 & 1.40 & 0.46 & & & 4.26 & 0.26\\
GRB081203A & 2.10 & 0.85 & 0.60 & & 1.00 \(^{5}\) & 5.60 & 0.35\\
GRB081221 & 2.26 & 6.90 & 1.64 & & & 6.00 & 0.31\\
GRB081222 & 2.77 & 0.51 & 0.62 & & 0.06 \(^{5}\) & 13.00 & 0.59\\
GRB090102* \(^{1,a}\) & 1.55 & 0.75 & 0.42 & & & 8.33 & 0.47\\
GRB090201 & 2.10 & 10.50 & 2.40 & & & 5.88 & 0.27\\
GRB090313 & 3.38 & 4.40 & 3.00 & & 0.20 \(^{5}\) & 3.67 & 0.35\\
GRB090328A & 0.74 & 0.65 & 0.82 & & & 2.36 & 0.45\\
GRB090404* \(^{2,a}\) & 3.00 & 12.10 & 3.20 & 0.20 \(^{2}\) & & 3.21 &
0.18\\
GRB090417B* \(^{2,a}\) & 0.34 & 3.30 & 0.64 & $-$1.90 \(^{2}\) & & 32.60 &
1.57\\
GRB090418A & 1.61 & 1.50 & 0.56 & & & 16.20 & 0.90\\
GRB090423 & 8.20 & 9.60 & 15.00 & & & 3.53 & 0.29\\
GRB090424 & 0.54 & 0.60 & 0.18 & & & 28.50 & 1.58\\
GRB090516 & 4.11 & 2.10 & 1.58 & & 0.54 \(^{5}\) & 2.40 & 0.11\\
GRB090530 & 1.27 & 0.39 & 0.34 & & & 3.54 & 0.28\\
GRB090618 & 0.54 & 0.24 & 0.07 & & & 24.90 & 0.71\\
GRB090709A & 1.80 & 2.20 & 0.70 & & & 17.70 & 0.79\\
GRB090715B & 3.00 & 1.30 & 1.22 & & 0.45 \(^{5}\) & 1.76 & 0.12\\
GRB090726 & 2.71 & 1.50 & 1.58 & & 0.63 \(^{5}\) & 0.64 & 0.06\\
GRB090809 & 2.74 & 0.96 & 1.24 & & 0.50 \(^{5}\) & 7.75 & 0.52\\
GRB090812* \(^{1,a}\) & 2.45 & 0.98 & 1.14 & & 2.00 \(^{5}\) & 6.54 &
0.59\\
GRB090902B & 1.82 & 2.30 & 1.30 & & & 3.27 & 0.28\\
GRB090926B* \(^{1,a}\) & 1.24 & 2.30 & 1.28 & & & 23.60 & 2.34\\
GRB091003 & 0.90 & 0.34 & 0.34 & & & 2.12 & 0.23\\
GRB091018 & 0.97 & 0.25 & 0.16 & & & 5.51 & 0.27\\
GRB091020 & 1.71 & 0.79 & 0.32 & & & 5.06 & 0.23\\
GRB091024 & 1.09 & 3.20 & 3.60 & & & 49.10 & 5.28\\
GRB091029 & 2.75 & 0.53 & 0.64 & & 0.05 \(^{5}\) & 5.74 & 0.29\\
GRB091109A & 3.08 & 1.30 & 1.98 & & & 1.93 & 0.20\\
GRB091127 & 0.49 & 0.10 & 0.12 & & & 80.10 & 5.29\\
GRB091208B & 1.06 & 1.10 & 0.50 & & & 7.61 & 0.55\\
GRB100424A & 2.46 & 3.40 & 2.60 & & & 35.50 & 3.77\\
GRB100615A* \(^{2,a}\) & 1.40 & 17.30 & 4.20 & $-$0.60 \(^{2}\) & & 24.90
& 1.49\\
GRB100621A & 0.54 & 2.80 & 0.56 & & & 17.00 & 0.81\\
GRB100728A & 1.57 & 2.60 & 0.78 & & & 33.30 & 1.31\\
GRB100816A & 0.80 & 0.24 & 0.32 & & & 2.17 & 0.23\\
GRB100901A & 1.41 & 0.34 & 0.28 & & & 11.70 & 0.54\\
GRB100906A & 1.73 & 0.60 & 0.92 & & & 3.11 & 0.16\\
GRB101219A & 0.72 & 0.81 & 0.82 & & & 5.88 & 0.91\\

\bottomrule
\end{tabularx}

\end{adjustwidth}
\end{table}

\begin{table}[H]\ContinuedFloat
\tablesize{\footnotesize}

   \caption{\textit{Cont}.}

\begin{adjustwidth}{-\extralength}{0cm}
\begin{tabularx}{\fulllength}{lCCccccc}
\toprule
\multirow{2}{*}{\textbf{GRB Name}} & \multirow{2}{*}{\textbf{Redshift}} & \boldmath{\(\NHXmath\)} & \boldmath{\(\NHXmath\)} \textbf{Error} & \multirow{2}{*}{\boldmath{\(\BOXmath\)}} &
\boldmath{\(\NHImath\)} & \textbf{Flux} & \textbf{Flux Error}\\
& & \boldmath{\(~10^{22}~{\rm cm}^{-2}\)} & \boldmath{\(~10^{22}~{\rm cm}^{-2}\)} & &
\boldmath{\(~10^{22}\NHIunit\)} & \boldmath{\(~10^{-12}~\fluxUnit\)} & \boldmath{\(~10^{-12}~\fluxUnit\)}
\\
\midrule

GRB101225A & 0.85 & 0.13 & 0.11 & & & 21.80 & 0.69\\
GRB110205A & 2.22 & 0.39 & 0.44 & & 0.28 \(^{5}\) & 4.65 & 0.23\\
GRB110422A & 1.77 & 1.70 & 0.52 & & & 23.30 & 1.00\\
GRB110503A & 1.61 & 0.24 & 0.22 & & & 6.19 & 0.25\\
GRB110715A & 0.82 & 1.50 & 0.84 & & & 11.60 & 0.59\\
GRB110818A & 3.36 & 1.40 & 1.46 & & 0.79 \(^{5}\) & 3.90 & 0.25\\
GRB111008A & 4.99 & 2.50 & 2.20 & & 2.51 \(^{5}\) & 8.43 & 0.39\\
GRB111117A & 2.21 & 1.50 & 2.40 & & & 1.18 & 0.19\\
GRB111209A & 0.68 & 0.22 & 0.10 & & & 10.40 & 0.44\\
GRB111228A & 0.71 & 0.35 & 0.13 & & & 9.18 & 0.41\\
GRB111229A & 1.38 & 0.57 & 0.78 & & & 5.64 & 0.50\\
GRB120118B & 2.94 & 5.70 & 3.40 & & & 3.99 & 0.35\\
GRB120119A & 1.73 & 0.98 & 0.92 & & 3.98 \(^{5}\) & 10.30 & 0.71\\
GRB120326A & 1.80 & 0.50 & 0.38 & & & 9.64 & 0.43\\
GRB120404A & 2.87 & 0.75 & 1.22 & & 0.05 \(^{5}\) & 7.60 & 0.57\\
GRB120711A & 1.41 & 2.00 & 0.56 & & & 16.80 & 0.67\\
GRB120712A & 4.17 & 1.50 & 2.40 & & 0.01 \(^{5}\) & 3.59 & 0.25\\
GRB120729A & 0.80 & 0.26 & 0.36 & & & 2.16 & 0.17\\
GRB120811C & 2.67 & 1.20 & 1.10 & & 0.32 \(^{5}\) & 7.24 & 0.53\\
GRB120815A & 2.36 & 0.71 & 1.24 & & 1.12 \(^{5}\) & 6.29 & 0.48\\
GRB120907A & 0.97 & 0.16 & 0.22 & & & 3.57 & 0.23\\
GRB120909A & 3.93 & 2.00 & 2.00 & & 0.50 \(^{5}\) & 9.58 & 0.49\\
GRB121024A & 2.30 & 1.10 & 1.02 & & 0.71 \(^{5}\) & 6.42 & 0.46\\
GRB121027A & 1.77 & 2.00 & 0.62 & & 6.31 \(^{5}\) & 11.20 & 0.61\\
GRB121209A & 2.10 & 12.10 & 3.40 & & & 4.34 & 0.26\\
GRB121211A & 1.02 & 0.78 & 0.32 & & & 3.11 & 0.21\\
GRB130420A & 1.30 & 0.40 & 0.26 & & & 5.22 & 0.31\\
GRB130427A & 0.34 & 0.11 & 0.08 & 0.95 & & 79.30 & 4.02\\
GRB130505A & 2.27 & 0.75 & 0.42 & & 0.04 \(^{5}\) & 56.60 & 2.03\\
GRB130511A & 1.30 & 0.75 & 0.58 & & & 1.80 & 0.19\\
GRB130603B & 0.36 & 0.44 & 0.18 & & & 3.94 & 0.30\\
GRB130702A & 0.14 & 0.05 & 0.08 & & & 11.40 & 0.88\\
GRB130907A & 1.24 & 1.10 & 0.20 & & & 53.20 & 1.58\\
GRB130925A & 0.35 & 3.10 & 0.58 & & & 75.30 & 3.64\\
GRB131004A & 0.72 & 0.66 & 0.54 & & & 4.71 & 0.50\\
GRB131030A & 1.29 & 0.42 & 0.32 & & & 10.30 & 0.64\\
GRB131103A & 0.60 & 1.40 & 0.54 & & & 5.76 & 0.45\\
GRB131105A & 1.69 & 2.00 & 0.88 & & & 3.92 & 0.27\\
GRB131108A & 2.40 & 0.96 & 1.40 & & 0.09 \(^{5}\) & 1.10 & 0.12\\
GRB131231A & 0.64 & 0.28 & 0.22 & & & 17.70 & 1.66\\
GRB140206A & 2.73 & 1.80 & 0.76 & & 0.32 \(^{5}\) & 21.70 & 0.79\\
GRB140213A & 1.21 & 0.14 & 0.24 & & & 13.20 & 0.52\\
GRB140301A & 1.42 & 0.85 & 0.88 & & & 1.90 & 0.27\\
GRB140304A & 5.28 & 4.10 & 4.80 & & & 25.30 & 1.69\\
GRB140419A & 3.96 & 1.20 & 1.04 & & 0.00 \(^{5}\) & 11.60 & 0.39\\
GRB140423A & 3.26 & 1.10 & 0.92 & & 0.03 \(^{5}\) & 8.49 & 0.43\\
GRB140430A & 1.60 & 0.74 & 0.94 & & 0.63 \(^{5}\) & 1.72 & 0.14\\
GRB140506A & 0.89 & 0.88 & 0.32 & & & 13.10 & 0.61\\
GRB140508A & 1.03 & 0.25 & 0.42 & & & 6.94 & 0.90\\
GRB140512A & 0.72 & 0.28 & 0.13 & & & 36.00 & 1.21\\
GRB140515A & 6.32 & 4.30 & 6.60 & & 0.00 \(^{5}\) & 4.44 & 0.33\\
GRB140518A & 4.71 & 2.90 & 3.20 & & 0.45 \(^{5}\) & 6.33 & 0.47\\
GRB140629A & 2.27 & 0.80 & 0.52 & & 1.00 \(^{5}\) & 3.11 & 0.16\\
GRB140703A & 3.14 & 1.00 & 1.60 & & 0.79 \(^{5}\) & 10.70 & 0.62\\
GRB140903A & 0.35 & 0.18 & 0.16 & & & 3.70 & 0.32\\
GRB140907A & 1.21 & 0.54 & 0.66 & & & 2.87 & 0.19\\
GRB141109A & 2.99 & 2.10 & 1.44 & & 1.26 \(^{5}\) & 8.30 & 0.56\\
GRB141121A & 1.47 & 0.42 & 0.42 & & & 11.50 & 0.83\\
GRB141220A & 1.32 & 0.38 & 0.54 & & & 3.09 & 0.35\\
GRB150206A & 2.09 & 1.10 & 0.44 & & 0.50 \(^{5}\) & 11.20 & 0.45\\
GRB150314A & 1.76 & 1.80 & 0.66 & & & 55.80 & 3.03\\
GRB150323A & 0.59 & 0.56 & 0.38 & & & 3.37 & 0.38\\
GRB150403A & 2.06 & 0.81 & 0.32 & & 0.63 \(^{5}\) & 28.40 & 0.82\\
GRB150821A & 0.76 & 2.20 & 0.70 & & & 8.24 & 0.60\\

\bottomrule
\end{tabularx}

\end{adjustwidth}
\end{table}

\begin{table}[H]\ContinuedFloat
\tablesize{\footnotesize}

   \caption{\textit{Cont}.}


\begin{adjustwidth}{-\extralength}{0cm}
\begin{tabularx}{\fulllength}{lCCccccc}
\toprule
\multirow{2}{*}{\textbf{GRB Name}} & \multirow{2}{*}{\textbf{Redshift}} & \boldmath{\(\NHXmath\)} & \boldmath{\(\NHXmath\)} \textbf{Error} & \multirow{2}{*}{\boldmath{\(\BOXmath\)}} &
\boldmath{\(\NHImath\)} & \textbf{Flux} & \textbf{Flux Error}\\
& & \boldmath{\(~10^{22}~{\rm cm}^{-2}\)} & \boldmath{\(~10^{22}~{\rm cm}^{-2}\)} & &
\boldmath{\(~10^{22}\NHIunit\)} & \boldmath{\(~10^{-12}~\fluxUnit\)} & \boldmath{\(~10^{-12}~\fluxUnit\)}
\\
\midrule
GRB151021A & 2.33 & 2.30 & 0.86 & & 1.58 \(^{5}\) & 11.60 & 0.64\\
GRB151027A & 0.81 & 0.43 & 0.13 & & & 34.90 & 1.32\\
GRB151027B & 4.06 & 3.30 & 4.20 & & 0.03 \(^{5}\) & 4.00 & 0.35\\
GRB151031A & 1.17 & 1.00 & 0.68 & & & 1.69 & 0.19\\
GRB160131A & 0.97 & 0.40 & 0.17 & & & 8.04 & 0.28\\
GRB160425A & 0.56 & 1.00 & 0.34 & & & 4.27 & 0.29\\
GRB160509A & 1.17 & 2.60 & 0.70 & & & 17.60 & 0.69\\
GRB160623A & 0.37 & 2.40 & 0.96 & & & 39.70 & 2.25\\
GRB160804A & 0.74 & 0.20 & 0.18 & & & 2.70 & 0.20\\
GRB161014A & 2.82 & 1.80 & 1.68 & & 0.25 \(^{5}\) & 17.10 & 1.17\\
GRB161117A & 1.55 & 0.82 & 0.36 & & & 14.00 & 0.67\\
GRB161219B & 0.15 & 0.20 & 0.05 & & & 38.90 & 1.34\\
GRB170113A & 1.97 & 0.92 & 0.58 & & & 16.50 & 0.76\\
GRB170519A & 0.82 & 0.37 & 0.17 & & & 8.11 & 0.44\\
GRB170705A & 2.01 & 1.20 & 0.40 & & & 19.20 & 0.74\\
GRB180205A & 1.41 & 0.50 & 0.56 & & & 8.95 & 0.80\\
GRB180325A & 2.25 & 1.30 & 0.76 & & 2.00 \(^{5}\) & 15.70 & 0.93\\
GRB180620B & 1.12 & 0.50 & 0.26 & & & 19.50 & 1.16\\
GRB180624A & 2.85 & 1.00 & 1.54 & & 3.16 \(^{5}\) & 2.29 & 0.21\\
GRB180720B & 0.65 & 0.34 & 0.11 & & & 207.00 & 7.86\\
GRB180728A & 0.12 & 0.07 & 0.06 & & & 41.20 & 1.06\\
GRB181010A & 1.39 & 2.30 & 0.64 & & & 8.17 & 0.42\\
GRB181020A & 2.94 & 0.75 & 0.62 & & 1.58 \(^{5}\) & 26.30 & 1.08\\
GRB190114A & 3.38 & 1.50 & 2.00 & & & 6.52 & 0.36\\
GRB190114C & 0.42 & 8.00 & 1.10 & & & 37.40 & 1.35\\
GRB190829A & 0.08 & 1.40 & 0.19 & & & 106.00 & 3.50\\
GRB191011A & 1.72 & 0.67 & 0.68 & & & 2.17 & 0.21\\
GRB201015A & 0.43 & 0.58 & 1.02 & & & 1.69 & 0.30\\
GRB201020B & 0.80 & 0.71 & 0.88 & & & 6.27 & 1.05\\
GRB201221A & 5.70 & 8.30 & 13.20 & & & 4.68 & 0.49\\
GRB210204A & 0.88 & 0.61 & 0.82 & & & 4.03 & 0.65\\
GRB210210A & 0.71 & 0.12 & 0.19 & & & 14.80 & 1.03\\
GRB210321A & 1.49 & 0.88 & 0.90 & & & 4.28 & 0.44\\
GRB210619B & 1.94 & 0.70 & 0.52 & & & 76.30 & 2.69\\
GRB210702A & 1.16 & 0.23 & 0.18 & & & 30.10 & 1.10\\
GRB211024B & 1.11 & 0.30 & 0.13 & & & 103.00 & 3.90\\
GRB211207A & 2.27 & 1.60 & 1.92 & & & 1.36 & 0.19\\
GRB221009A & 0.15 & 1.40 & 0.76 & & & 229.00 & 17.90\\
GRB230325A & 1.66 & 1.70 & 1.22 & & & 2.71 & 0.34\\
GRB231118A & 0.83 & 0.89 & 0.44 & & & 45.00 & 4.00\\
GRB240419A & 5.18 & 11.30 & 13.20 & & & 2.20 & 0.33\\

\bottomrule
\end{tabularx}

\end{adjustwidth}
\footnotesize{{\textsuperscript{1}} 
\textls[-15]{\citet{greiner2011}; \textsuperscript{2} \citet{chrimes2019}; \textsuperscript{3}\citet{fynbo2009}; \textsuperscript{4} \citet{hashimoto2010}; \textsuperscript{5} \citet{tanvir2019}; \textsuperscript{6} \citet{perley2009}}. Auth{The [*] after GRBs name indicate optically dark GRB. } Auth{Indicators for} Darkness Classification Method : {a.} 
 Jakobsson b. van der Horst.}
\end{table}





}

\normalsize

\vspace{6pt}




\authorcontributions{{Eka Puspita Arumaningtyas as E.P.A., Hasan
Al-Rasyid as H.A.R., Maria Giovanna Dainotti as M.G.D., and Daisuke
Yonetoku as D.Y. E.P.A. and D.Y.:} 
 conceptualization; E.P.A., M.G.D., and
D.Y.: methodology; M.G.D. and D.Y.: validation; E.P.A., M.G.D., and
D.Y.: formal analysis; E.P.A. and H.A.R.: data curation; E.P.A.: writing---original draft preparation; E.P.A., H.A.R., M.G.D., and D.Y.: writing---review and editing; E.P.A. and H.A.R.: software implementation and
visualization; M.G.D. and D.Y.: supervision; D.Y.: funding acquisition.
All authors have read and agreed to the published version of the
manuscript.}

\funding{This research was supported by Project for Remarkable
Ph.D.~Students in Next Generation of Kanazawa University, JSPS KAKENHI
Grant Numbers 23H04898 (DY), 21H01090 (DY), 20K20525 (DY), and Sakigake
project 2022 of Kanazawa University.}

\dataavailability{We analyzed the spectral properties of GRBs from the
SWIFT-XRT GRB Catalogue (\url{https://www.swift.ac.uk/xrt\_live\_cat/} ({accessed on 16 August 2024}.
)).}




\conflictsofinterest{The authors declare no conflicts of interest. The
funders had no role in the design of the study; in the collection,
analyses, or interpretation of data; in the writing of the manuscript;
or in the decision to publish the results.}

\begin{adjustwidth}{-\extralength}{0cm}

\reftitle{References}

\PublishersNote{}
\end{adjustwidth}
\end{document}